\documentclass{aa}  

\usepackage{graphicx}
\usepackage{txfonts}
\usepackage{lipsum}
\usepackage{subcaption}         
\usepackage{lscape}             
\usepackage{placeins}           
\usepackage[
colorlinks=true,
linkcolor=blue,
citecolor=blue,
urlcolor=blue
]{hyperref}

\begin{document}

\title{The Influence of Evaporation on the Formation and Evolution of Huntsman Systems}

\author{Hao-Ran Yang\inst{1}\corrauth{yanghaoran@tyut.edu.cn}      
\and Shi-Jie Gao\inst{2,3}
\and Xiang-Dong Li\inst{2,3}
\and Yun-Lang Guo\inst{2,3}
\and Wen-Shi Tang\inst{4}
}
\institute{College of Physics and Optoelectronics, Taiyuan University of Technology, Taiyuan 030024, China
\and School of Astronomy and Space Science, Nanjing University, Nanjing 210023, China
\and Key Laboratory of Modern Astronomy and Astrophysics (Nanjing University), Ministry of Education, Nanjing 210023, China
\and Department of Astronomy, Xiamen University, Xiamen 361005, China}

\date{Received XXX, 20XX}

\abstract
{Huntsman systems are a recently identified and rare subclass of millisecond pulsar (MSP) binaries, characterized by a detached neutron star and an evolved giant companion in relatively wide orbits. Their formation has been proposed to involve red-bump-induced detachment, whereas the influence of MSP-driven evaporation during this evolutionary stage has not yet been quantitatively assessed.} 
{We quantify the influence of MSP-driven evaporation on the formation and evolution of Huntsman systems and assess its effects on their observable properties.}
{We performed detailed binary evolution calculations including MSP-driven evaporation over a wide range of initial binary parameters and combined them with binary population synthesis to predict the observable population of Huntsman systems.}
{We find that Huntsman systems originate primarily from binaries undergoing Case B mass transfer and red-bump-induced detachment, with initial donor masses of $1.0-2.5\ M_\odot$ and orbital periods above the bifurcation period. Evaporation has a secondary effect, slightly modifying the orbital evolution and the duration of the detached phase, but does not significantly alter the formation parameter space or the expected population, which contains about 10 Huntsman systems. We further show that evaporation can produce systematic shifts in the white dwarf mass-orbital period relation at the low-mass end, leading to systematically wider final orbits for a given white dwarf mass.}
{}


\maketitle
\nolinenumbers
\section{Introduction}
Huntsman systems are a newly identified and extremely rare subclass of millisecond pulsar (MSP) binaries, characterized by a fully recycled neutron star (NS) orbiting a heavily stripped red-giant companion in a binary with a relatively long orbital period (typically several to ten days) \citep{2015ApJ...804L..12S,2025ApJ...980..124S}. Unlike the compact “spider” binaries \citep{2013IAUS..291..127R}, where the pulsar wind in a tight orbit ablates a non-degenerate or semi-degenerate companion star \citep{2025ApJ...994....8K}, Huntsman binaries are detached systems in which the evolved companion underfills its Roche lobe \citep{2025ApJ...980..124S}. Only two confirmed members, PSR J1417-4402 \citep{2015ApJ...804L..12S,2018ApJ...866...83S} and PSR J1947-1120 \citep{2025ApJ...980..124S}, have been identified so far through multi-wavelength observations, revealing radio pulsations, $\gamma$-ray emission, and optical signatures consistent with evolved giant companions.

Theoretically, Huntsman pulsars are believed to form near the “red-bump” phase \citep{1967ZA.....67..420T,2015MNRAS.453..666C} of stellar evolution, when a temporary contraction of the red-giant envelope interrupts Roche-lobe overflow (RLO), leaving a detached, partially stripped donor \citep{2025ApJ...980..124S,2025A&A...698L...5B}. Stellar evolution modeling supports this scenario, suggesting that Huntsman systems represent a short-lived post-accretion stage linking low-mass X-ray binaries (LMXBs) and wide MSP binaries \citep{2025A&A...698L...5B}. On the other hand, \citet{2025PASJ..tmp...65L} proposed that the observed properties of Huntsman pulsar binaries could be explained by assuming a normal red-giant companion rather than a red-bump star, when possible irradiation effects are considered. While these scenarios explain the observed detachment and long orbital periods, the potential impact of MSP-driven evaporation on the formation and evolutionary pathways of Huntsman systems remains uncertain.

MSP-driven evaporation is a well-established mechanism in compact interacting binaries, where the rotational energy carried by the pulsar wind irradiates and ablates the companion star \citep{1988Natur.334..227V,1989ApJ...336..507R,2014ApJ...786L...7B}. This process is responsible for the pronounced mass loss observed in black widows and redbacks, and plays a central role in shaping their orbital evolution and donor properties \citep{2013ApJ...775...27C,2015ApJ...814...74J,2022MNRAS.515.2725G}. Whether a similar mechanism operates in Huntsman systems, however, remains uncertain. Although their wider orbits reduce the pulsar-wind flux at the companion surface, the larger size of the evolved companion can partly compensate for this effect in terms of the solid angle it subtends as viewed from the pulsar. Moreover, the extended and relatively low-density envelopes of evolved giant donors may also be susceptible to MSP-driven mass loss. The efficiency and evolutionary consequences of evaporation in Huntsman systems therefore remain to be quantified. In particular, it remains unclear whether evaporation can modify the duration of the Huntsman phase, further strip the donor after detachment, or the properties of the final white dwarf (WD) remnant. It also remains unclear to what extent evaporation affects the expected number and observable distribution of Huntsman systems. Clarifying the efficiency of MSP-driven evaporation during or after the red-bump phase is therefore important for quantitatively assessing its role in the formation and subsequent evolution of Huntsman systems.

In this paper, we incorporate MSP-driven evaporation into detailed binary evolution calculations and assess its impact on the formation and subsequent evolution of Huntsman systems, as well as on their expected population properties. The rest of the paper is organized as follows. In Section~\ref{s:method}, we introduce the binary evolution models used in this paper, including the MSP-driven evaporation models and other prescriptions. The simulated results are presented in Section~\ref{s:result} and compared with observations. Our discussion and conclusions are given in Section~\ref{s:conclusion}.

\section{Binary evolution model}\label{s:method}

Huntsman pulsars are thought to originate from NS LMXBs, and we therefore model their formation and evolution within the standard evolutionary framework of NS LMXBs. We carried out detailed binary evolution calculations using the binary evolution code MESA (version number r22.11.1; \citealt{2011ApJS..192....3P,2013ApJS..208....4P,2015ApJS..220...15P,2018ApJS..234...34P,2019ApJS..243...10P,2023ApJS..265...15J}). The NS in the LMXBs is treated as a mass point with an initial mass of $M_{\rm NS}=1.4\ M_{\odot}$. The donor star is initially assumed to be a main-sequence star with mass $M_{\rm d,0}\in [0.9,2.7]\ M_{\odot}$ and orbital period $\log P_{\rm orb,0}\ (\rm d)\in [-0.2,2.2]$, using a grid spacing of 0.1 $M_{\odot}$ in donor mass and 0.1 in $\log P_0$. We adopt a fixed initial donor metallicity of $Z=0.02$ and the \citet{1988A&A...202...93R} scheme to calculate the mass transfer (MT) rates $\dot{M}_{\rm tr}$ via RLO. The mass accretion efficiency onto the NS is set to 0.3 \citep{1999A&A...350..928T}, and the accretion rate $\dot{M}_{\rm acc}$ of the NS is also limited by the Eddington accretion rate. We also include angular-momentum loss due to magnetic braking and gravitational wave radiation. All the systems are evolved until the Hubble time unless the calculations fail to converge. 

We assume that evaporation becomes effective only after RLO has effectively ceased and the NS has entered the radio-ejection phase. In practice, this condition is implemented by requiring the mass-transfer rate to fall below $\sim 10^{-20}\ M_{\odot} \ \rm yr^{-1}$, which is used here as a numerical threshold rather than a physically derived critical rate. The corresponding evaporation-driven mass-loss rate from the donor is written as follows \citep{1992MNRAS.254P..19S}:
\begin{equation}
  \dot{M}_{\rm d,evap}=-\frac{f}{2v_{\rm d, esc}^2}L_{\rm P}\left(\frac{R_{\rm d}}{a}\right)^2 \label{eq:evap},
\end{equation}
where $R_{\rm d}$, $v_{\rm d, esc}$ and $a$ denote the donor radius, the escape velocity at the donor surface, and the binary orbital separation, respectively. Here $f$ is a free parameter describing the evaporation efficiency \citep{1989ApJ...336..507R,1989ApJ...343..292R,2020MNRAS.495.3656G}. We consider $f\leq 0.1$ in this work, consistent with previous studies \citep{2013ApJ...775...27C,2015ApJ...814...74J}. The pulsar spin-down luminosity is given by $L_{\rm P}=4\pi^2I\dot{P}_{\rm s}/P_{\rm s}^3$, where $I$, $P_{\rm s}$ and $\dot{P}_{\rm s}$ are the moment of inertia, the spin period and the spin-period derivative of the NS, respectively. Under the standard magnetic-dipole radiation model, $\dot{P}_{\rm s}$ is approximated as $\dot{P}_{\rm s}=4\pi^2\mu^2/(Ic^3P_{\rm s})$ \citep{2023ApJ...945....2Y}, where $\mu$ is the magnetic dipole moment of the NS. We adopt $I=10^{45}\ \rm g\ cm^2$, $\mu=10^{26}\ \rm G\ cm^3$ and the initial spin period $P_{\rm s,0}=2\ \rm ms$. The pulsar spin evolution is then followed assuming pure magnetic-dipole braking.

\section{Results}\label{s:result}

\begin{figure*}
\sidecaption
  \includegraphics[width=12cm]{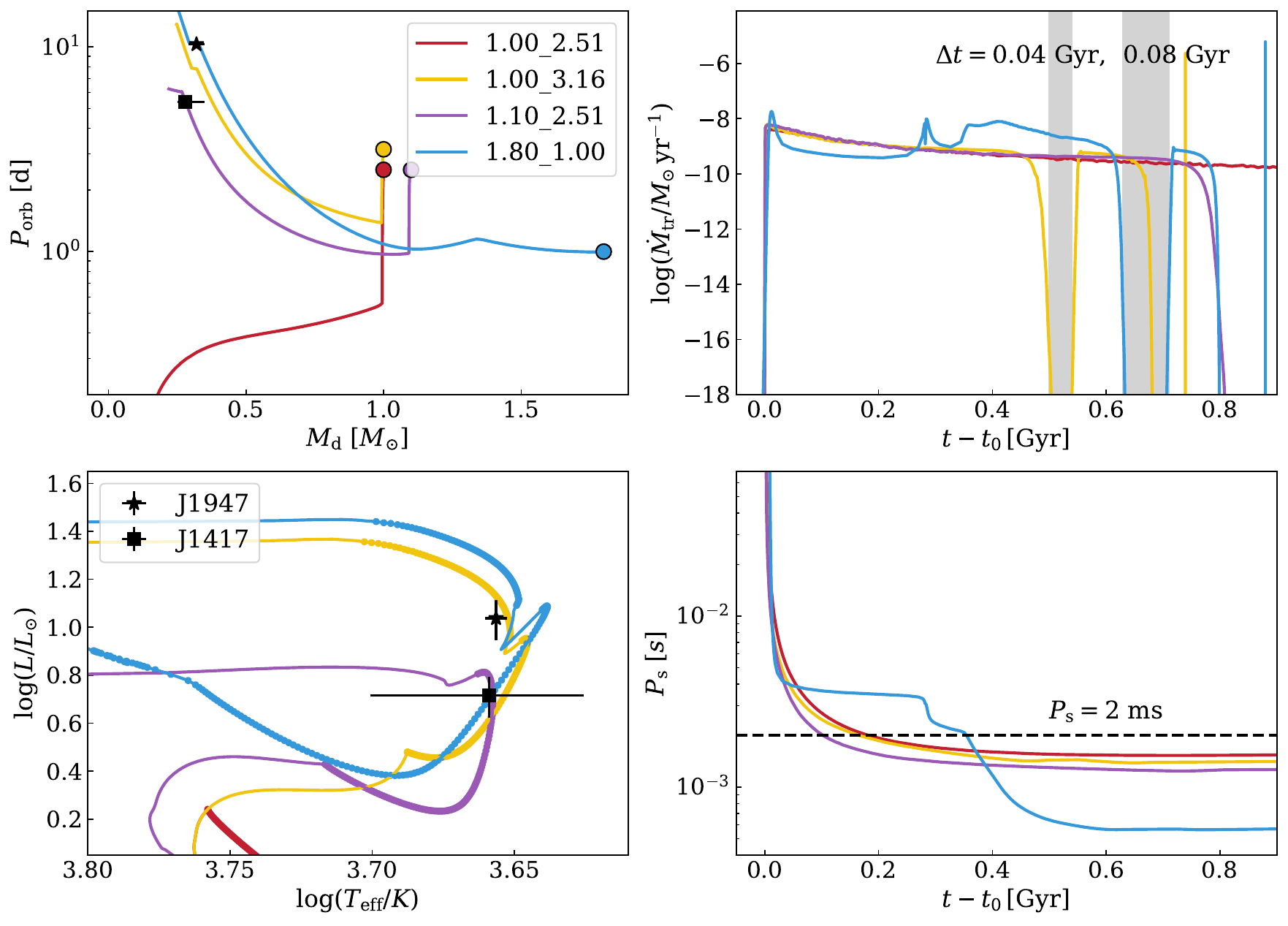}
	\caption{Example evolution for systems with different initial parameters under an evaporation efficiency of $f=0.01$. Different colors denote different initial combinations of donor mass $M_{\rm d,0}$ and orbital period $P_{\rm orb,0}$. The left panels show the evolution in the $M_{\rm d}-P_{\rm orb}$ plane and the Hertzsprung-Russell (H-R) diagram, respectively, where the star and square symbols indicate the two confirmed Huntsman systems, J1947 and J1417. And the circles in the $M_{\rm d}-P_{\rm orb}$ plane mark the starting points of evolution. The right panels present the evolution of the mass transfer (MT) rate $\dot{M}_{\rm tr}$ and the pulsar spin period $P_{\rm s}$. The time axis is shifted such that $t_0$ corresponds to the onset of Roche-lobe overflow (RLO). The dotted segments in the H-R diagram indicate phases during which the system is undergoing MT. And the gray regions in the upper right panel represent phases when the MT is interrupted.}
	\label{fig:example}
\end{figure*}
\begin{figure*}
\sidecaption
  \includegraphics[width=12cm]{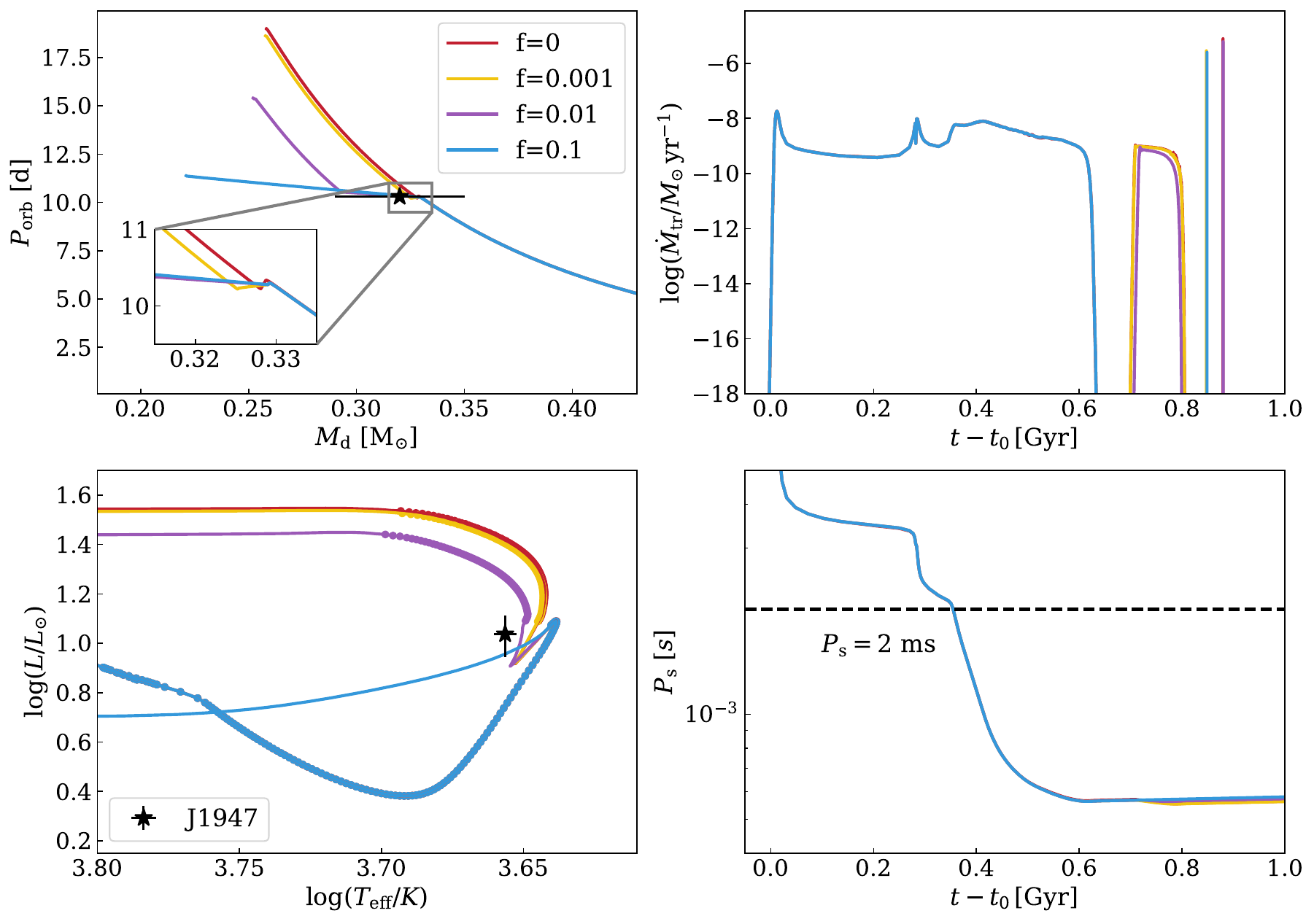}
	\caption{Effect of different evaporation efficiencies on the binary evolution for a system with initial parameters $(M_{\rm d,0}, P_{\rm orb,0}) = (1.80\,M_\odot, 1.00\,{\rm d})$. The evaporation efficiency is taken to be $f = 0$, $0.001$, $0.01$, and $0.1$, denoted by different colors.}
	\label{fig:f_example}
\end{figure*}
\begin{figure*}
\sidecaption
  \includegraphics[width=12cm]{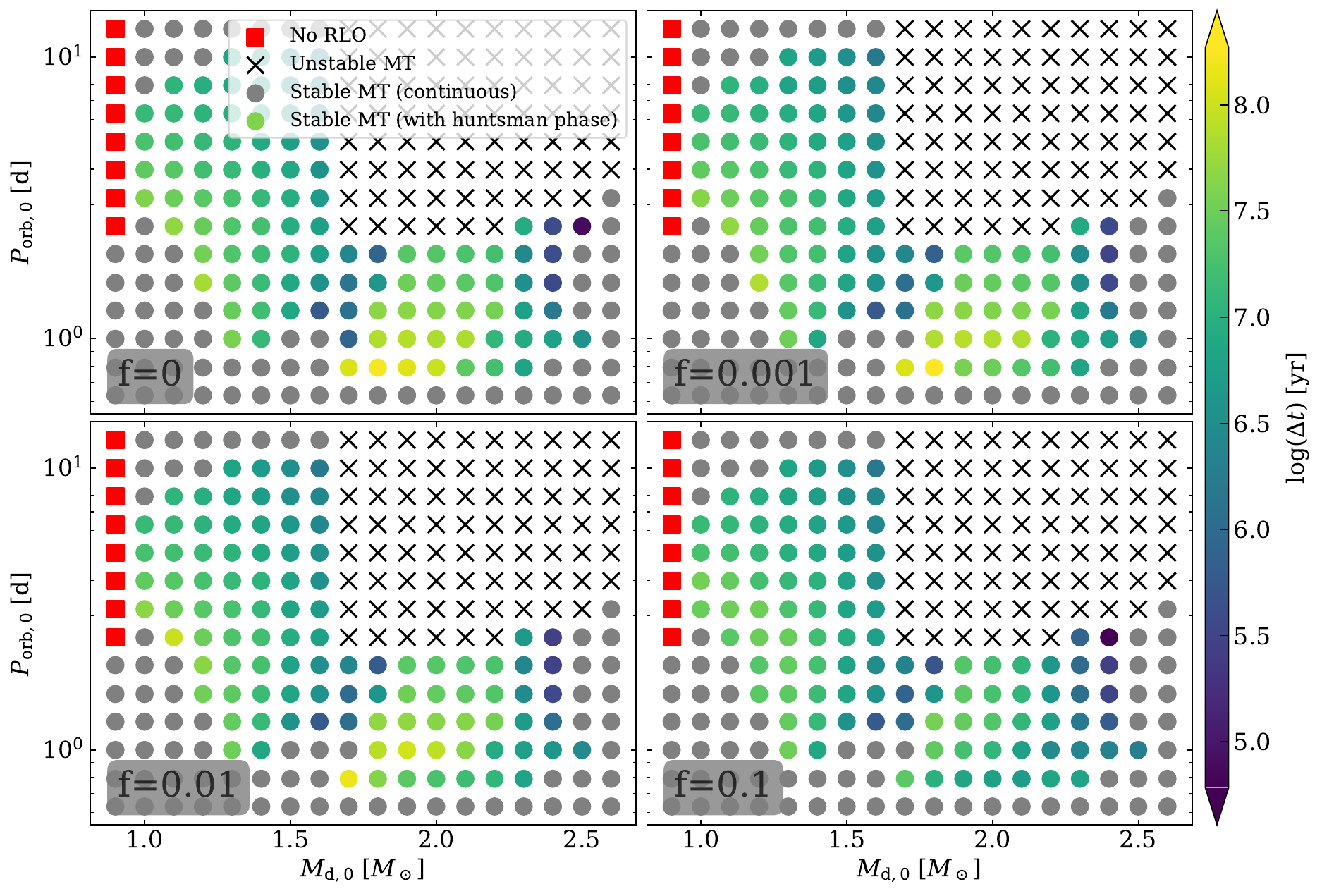}
	\caption{Parameter space for the formation of Huntsman systems under different evaporation efficiencies. The horizontal and vertical axes represent $M_{\rm d,0}$ and $P_{\rm orb,0}$, respectively. Red squares denote systems that do not undergo RLO, while black crosses indicate systems with dynamically unstable MT. Gray and colored circles correspond to systems with continuously stable MT and systems in which MT is interrupted, respectively. The color encodes the duration of the Huntsman phase.}
	\label{fig:space}
\end{figure*}
\begin{figure*}
\sidecaption
  \includegraphics[width=12cm]{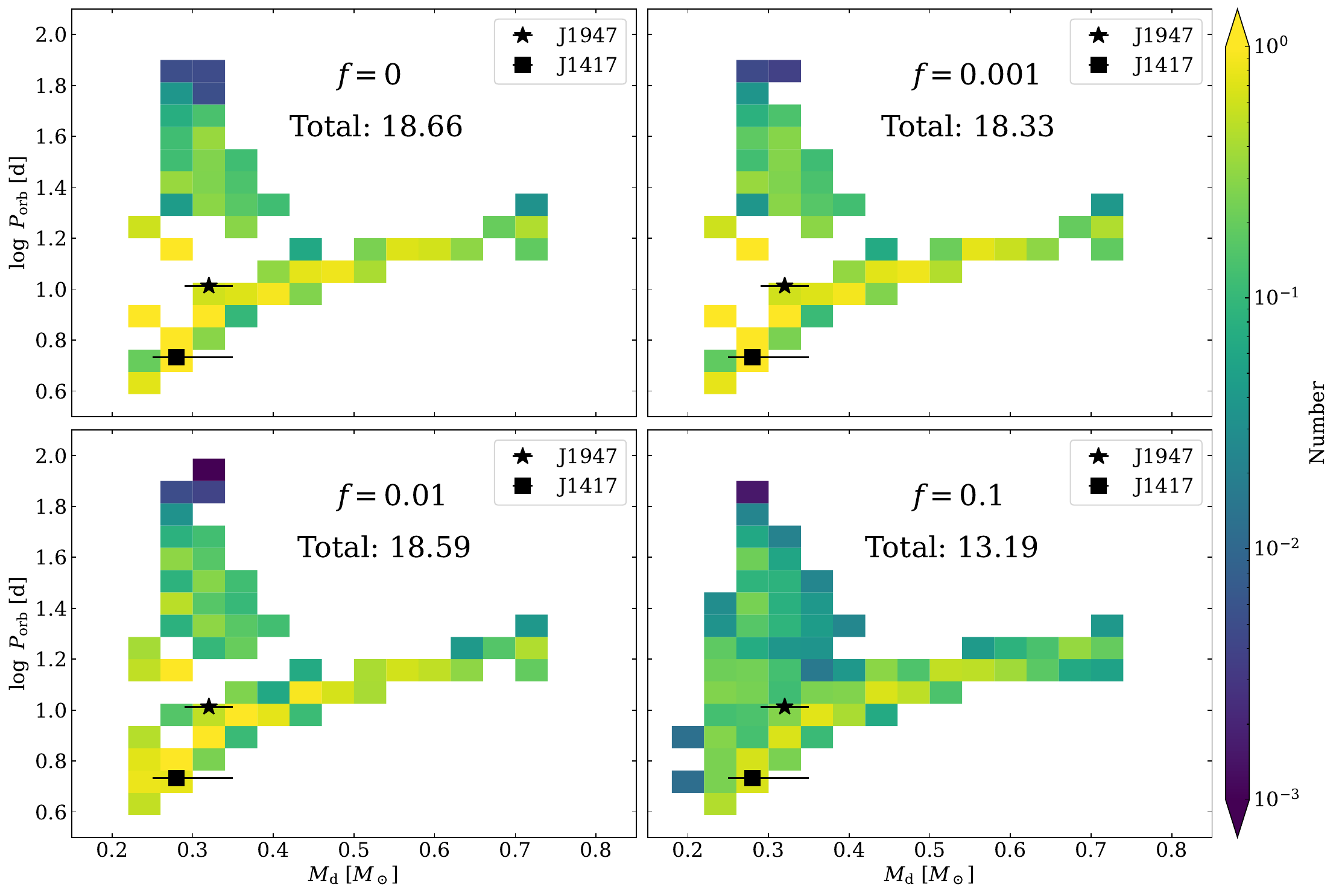}
	\caption{Predicted distribution of Huntsman systems in the observable $M_{\rm d}-P_{\rm orb}$ plane under different evaporation efficiencies. The color scale represents the number density of systems in each bin. The population synthesis is based on a total of $10^6$ binary systems.}
	\label{fig:pop}
\end{figure*}

In this section, we present the main results of our calculations and compare them with the currently known Huntsman systems. The key observational properties of the known Huntsman systems are first summarized for comparison with our models. We then present representative evolutionary tracks and discuss the effect of evaporation on their evolution. Finally, we explore the initial parameter space for forming Huntsman systems and the corresponding population predictions.

\subsection{Observed Huntsman systems}\label{s:obs}

\begin{table}[t]
\caption{Observed properties of the two confirmed Huntsman systems.}
\label{tab:Huntsman_obs}
\centering
\begin{tabular}{lcc}
\hline\hline
Property & J1417 & J1947 \\
\hline
$P_{\rm orb}$ (d) & 5.374 & 10.265 \\
$M_{\rm d}$ ($M_\odot$) & $0.28^{+0.07}_{-0.03}$ & $0.32 \pm 0.03$ \\
$R_{\rm d}$ ($R_\odot$) & $3.7 \pm 0.3$ & $5.4 \pm 0.3$ \\
$L_{\rm d}$ ($L_\odot$) & $5.2 \pm 1.0$ & $10.9 \pm 2.1$ \\
$T_{\rm d}$ (K) & $4560^{+460}_{-336}$ & $4534 \pm 41$ \\
$f_2$ & $0.83^{+0.05}_{-0.07}$ & $0.87 \pm 0.02$ \\
Mass Ratio ($M_{\rm d}/M_{\rm NS}$) & $0.171 \pm 0.002$ & $0.182 \pm 0.001$ \\
Gaia distance (kpc) & $4.2^{+1.0}_{-0.7}$ & $5.7^{+2.0}_{-1.3}$ \\
$P_{\rm s}$ (ms) & 2.664 & 2.240 \\
\hline
\end{tabular}

\tablefoot{
The properties of J1417 and J1947 are compiled from the observational analyses of \citet{2016ApJ...820....6C}, \citet{2018ApJ...866...83S} and \citet{2025ApJ...980..124S}.}

\end{table}

At present, the observational sample of Huntsman systems remains extremely limited. The first confirmed member is PSR J1417-4402 (hereafter J1417) \citep{2015ApJ...804L..12S,2018ApJ...866...83S}, a fully recycled MSP with a spin period of 2.66 ms \citep{2016ApJ...820....6C} in a 5.374 d orbit around a heavily stripped giant companion of mass $\sim 0.28\,M_\odot$. The second confirmed member is PSR J1947-1120 (hereafter J1947) \citep{2025ApJ...980..124S}, which has a spin period of 2.24 ms and orbits a red-giant companion of mass $\sim 0.32\,M_\odot$ in a binary with an orbital period of 10.265 d. The main observed properties of these two systems are summarized in Table~\ref{tab:Huntsman_obs}. In addition to the orbital period $P_{\rm orb}$ and donor mass $M_{\rm d}$ mentioned above, the table also lists the donor radius $R_{\rm d}$, luminosity $L_{\rm d}$, effective temperature $T_{\rm d}$, Roche-lobe filling factor $f_2$, mass ratio, Gaia distance and spin period $P_{\rm s}$ of the NS. We also note that 2FGL J0846.0+2820 \citep{2017ApJ...851...31S} has been discussed as a Huntsman candidate, with an 8.1 d orbit and a partially stripped red-giant companion, although no MSP has yet been confirmed in this source. These systems indicate that Huntsman pulsars are characterized by relatively long orbital periods, fully recycled pulsars, and giant companions in detached configurations, and they therefore provide the main observational benchmarks for the evolutionary models presented below.

\subsection{Representative evolutionary tracks}\label{s:case}

  Figure~\ref{fig:example} presents representative evolutionary tracks of binary systems with different initial parameters under an evaporation efficiency of $f=0.01$. Different colors correspond to different initial combinations of $M_{\rm d,0}$ and $P_{\rm orb,0}$. The left panels show the evolution of the systems in the $M_{\rm d}-P_{\rm orb}$ plane and the Hertzsprung-Russell (H-R) diagram, respectively, while the right panels illustrate the temporal evolution of $\dot{M}_{\rm tr}$ and $P_{\rm s}$. The dotted segments in the H-R diagram indicate phases during which the system is undergoing MT. And the time axes in the right panels are shifted such that $t_0$ corresponds to the onset of RLO. For reference, the positions of the two confirmed Huntsman systems, J1947 and J1417, are also marked in the parameter space. 

In Figure~\ref{fig:example}, the red and yellow solid lines both correspond to systems with an initial donor mass of $M_{\rm d,0}=1.0\,M_\odot$, but with different initial orbital periods of $P_{\rm orb,0}=2.51$ d and $3.16$ d, respectively. For the system with $P_{\rm orb,0}=2.51$ d, MT begins at an early evolutionary stage of the donor (Case A MT\footnote{Case A, B and C MT refer to the RLO process when the donor is undergoing core hydrogen burning, shell hydrogen burning and stages after exhaustion of the core helium burning, respectively.}). The binary orbit subsequently shrinks, and the system does not experience the red-bump phase. As a result, it cannot evolve into a Huntsman system. In contrast, for the system with $P_{\rm orb,0}=3.16$ d, longer than the bifurcation period which separates the converging from diverging binary systems \citep{1988A&A...191...57P,1989A&A...208...52P}, MT starts later and proceeds when the donor has evolved to the giant branch (Case B MT). When the donor reaches the red-bump phase, its temporary contraction causes it to detach from its Roche lobe, and the MT is interrupted. This phase corresponds to the hook-like feature in the H-R diagram, as well as the plateau in the $M_{\rm d}-P_{\rm orb}$ plane where $P_{\rm orb}$ remains nearly constant or increases slowly. This detached stage is identified as the observational phase of Huntsman systems. Afterward, the donor expands again, refills its Roche lobe, and MT resumes. These results indicate that only systems with initial orbital periods above the bifurcation period are able to evolve through the Huntsman phase. Additionally, we further present two evolutionary tracks with $(M_{\rm d,0}, P_{\rm orb,0}) = (1.1\,M_\odot, 2.51\,{\rm d})$ and $(1.8\,M_\odot, 1.0\,{\rm d})$, which can approximately reproduce the observed properties of J1947 and J1417, respectively. For the system with $(M_{\rm d,0}, P_{\rm orb,0}) = (1.1\,M_\odot, 2.51\,{\rm d})$, the donor does not refill its Roche lobe after passing through the red-bump phase. From the evolution of $\dot{M}_{\rm tr}$, the duration of the Huntsman phase is found to be on the order of several tens of Myr. We also follow the spin evolution of the NS and find that during the Huntsman phase the pulsar spin period can reach the millisecond range, consistent with current observations. We note that the appearance of sub-millisecond pulsars in some models indicates that additional spin-down mechanisms, not included in the present calculations, may become important near the spin equilibrium. 

Figure~\ref{fig:f_example} illustrates the effect of different evaporation efficiencies on the evolution of Huntsman systems. The initial binary parameters are fixed at $(M_{\rm d,0}, P_{\rm orb,0}) = (1.80\,M_\odot, 1.00\,{\rm d})$, while the evaporation efficiency is varied over $f = 0$, $0.001$, $0.01$, and $0.1$. From the upper-left panel, it can be seen that in the absence of evaporation ($f=0$, red solid line), $P_{\rm orb}$ decreases abruptly when the donor temporarily detaches from its Roche lobe during the red-bump phase, and then increases again after the donor refills its Roche lobe. When evaporation is included, however, the evolution of $P_{\rm orb}$ during this phase becomes significantly smoother, and in some cases even shows a gradual increase. The upper-right panel suggests that increasing $f$ may slightly prolong the duration of the detached phase. For sufficiently large evaporation efficiency ($f=0.1$, blue solid line), the mass loss driven by the MSP becomes strong enough that the donor is unable to refill its Roche lobe after detachment, except for occasional MT episodes induced by H-flashes. We also find that different values of $f$ have only a minor impact on the spin evolution of the NS, and $P_{\rm s}$ remains in the millisecond range during the Huntsman phase.

\subsection{Formation space and population predictions}\label{s:population}

Figure~\ref{fig:space} shows the initial parameter space $M_{\rm d,0}-P_{\rm orb,0}$ for the formation of Huntsman systems under different evaporation efficiencies. In this work, we define the Huntsman phase as the evolutionary stage during which the system is detached (i.e., no ongoing MT), the NS has been spun up to the millisecond regime (with an accreted mass larger than $0.1\ M_\odot$), and the donor star is in the giant phase. Systems that do not initiate RLO are marked by red squares, while those experiencing dynamically unstable MT are indicated by black crosses. Systems undergoing stable mass transfer are shown as circles. Among them, gray circles represent binaries with continuous MT, whereas colored circles correspond to systems in which the MT is interrupted, i.e., those that can evolve through a Huntsman phase. The color further encodes the duration of the Huntsman phase. From Figure~\ref{fig:space}, it can be seen that for systems with $M_{\rm d,0} < 1.0\,M_\odot$, the binaries either undergo continuous MT or do not initiate MT at all. In contrast, for $M_{\rm d,0} > 2.5\ M_\odot$, the systems tend to exhibit either stable continuous MT or dynamically unstable MT. As a result, the formation of Huntsman systems is restricted to an intermediate donor mass range of $M_{\rm d,0} \simeq 1.0-2.5\ M_\odot$. In addition, $P_{\rm orb,0}$ should be larger than the bifurcation period corresponding to a given donor mass, but not excessively large. In our calculations, the upper limit is typically $\sim 10$ d and becomes smaller (approximately $2-3$ d) for systems with more massive donors. We find that the overall distribution of the parameter space leading to Huntsman systems is not significantly affected by the evaporation efficiency. The duration of the Huntsman phase spans a wide range from $\sim 10^5$ to $10^8$ yr. There is a weak tendency for the duration to decrease with increasing $f$, which may be attributed to enhanced mass loss driven by the pulsar, causing the donor to leave the giant phase at an earlier stage. This suggests that the occurrence of Huntsman systems is primarily determined by the intrinsic binary evolution (i.e., the red-bump-induced detachment), while evaporation plays a secondary role.

Our results are broadly consistent with previous evolutionary studies that associate Huntsman systems with red-bump-induced detachment \citep{2025ApJ...980..124S,2025A&A...698L...5B}. \citet{2025PASJ..tmp...65L} also explored an alternative scenario in which irradiation allows the observed properties of Huntsman systems to be reproduced with a normal red-giant companion. In this work, we additionally investigate the role of MSP-driven evaporation in the formation and evolution of Huntsman systems, finding that it is not required for their formation but can modestly modify their subsequent evolution.

We also performed binary population synthesis using the BSE code \citep{Hurley+2002+BSE} to estimate the birth rate of the progenitors of LMXBs. The simulations assumed a constant star formation rate of $3~M_\odot~{\rm yr^{-1}}$ within the past $12~{\rm Gyr}$, a stochastic supernova mechanism \citep{Mandel+2020}, a common-envelope efficiency parameter $\alpha_{\rm CE} = 1$, and the Nanjing $\lambda$ prescription \citep{Xu+2010}. The primary masses of primordial binaries were drawn from a Kroupa initial mass function \citep{Kroupa+2001} in the range of $3-50~M_\odot$, while the mass ratios were sampled from a flat distribution between 0 and 1. The initial orbital separations were assumed to be uniformly distributed in logarithmic space between 3 and $10^{5}~{R_\odot}$.  In total, $10^7$ binary systems were evolved to estimate the birth rate. The birth rate was then multiplied by the duration of the corresponding evolutionary phase to estimate the simulated number of Huntsman systems. Figure~\ref{fig:pop} presents the results of the population synthesis calculations. The simulations suggest that the number of Huntsman systems is $\sim 10$ under the adopted population synthesis prescriptions. We further find that the number of systems decreases slightly with increasing evaporation efficiency, from 18.66 for $f=0$ to 13.19 for $f=0.1$. At the same time, the distribution of Huntsman systems in the $M_{\rm d}-P_{\rm orb}$ plane appears to broaden modestly when stronger evaporation is included.

\section{Discussion and conclusions}\label{s:conclusion}

\subsection{The effect of evaporation on the white dwarf mass-orbital period relation}

\begin{figure}[t]
	\centering \includegraphics[width=0.48\textwidth]{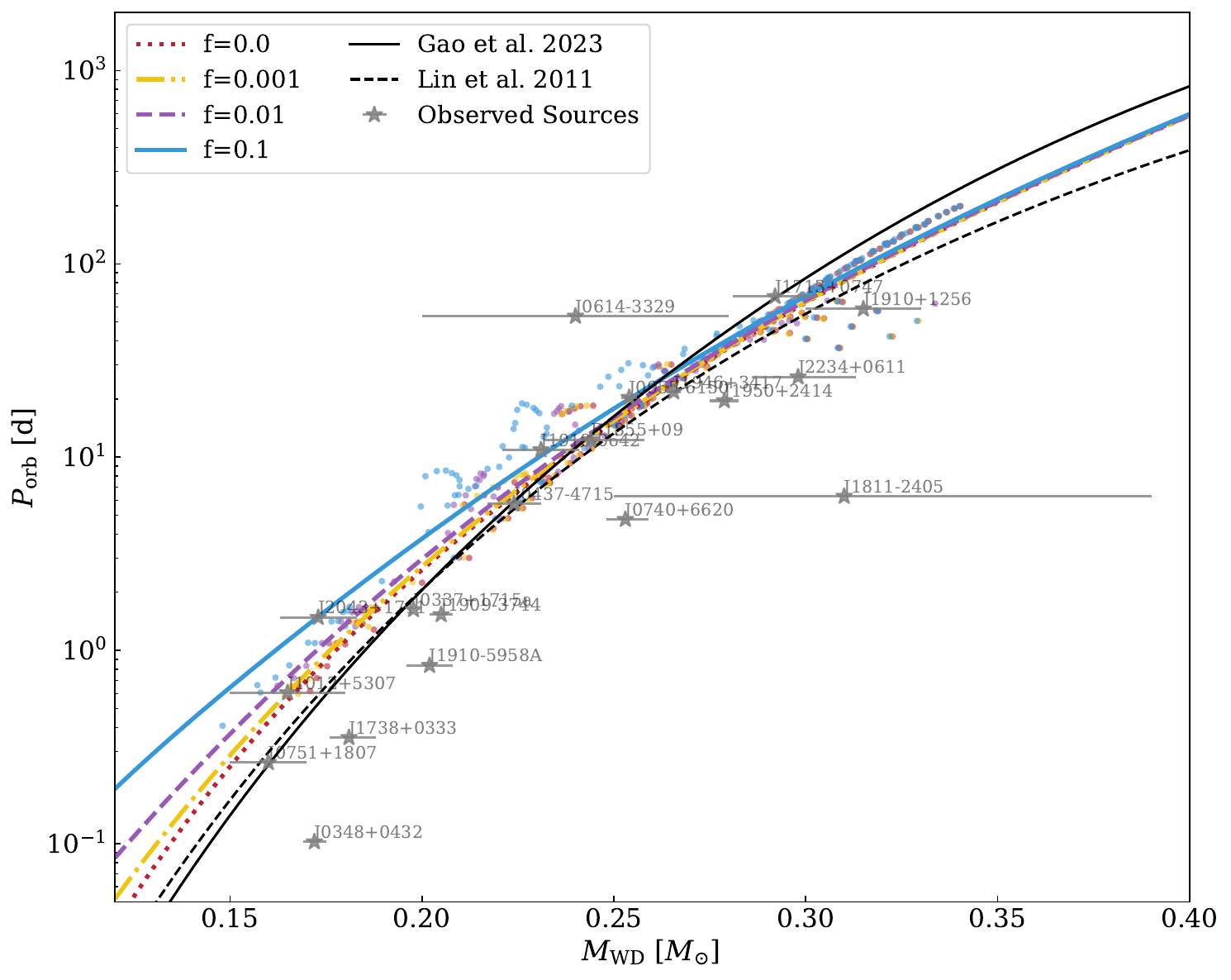}
	\caption{The orbital period $P_{\mathrm{orb}}$ versus white dwarf mass $M_{\mathrm{WD}}$ for various evaporation efficiencies $f$. The colored lines represent our power-law-corrected fits for different efficiencies: $f=0.0$ (red dotted line), $f=0.001$ (yellow dash-dotted line), $f=0.01$ (purple dashed line), and $f=0.1$ (blue solid line). The underlying semi-transparent scatter points denote the corresponding binary evolution simulation data. For comparison, the theoretical models of \citet{2011ApJ...732...70L} (black dashed line) and \citet{2023MNRAS.525.2605G} (black solid line) are shown. Gray stars with error bars represent observed binary millisecond pulsar systems with helium white dwarf companions.}
	\label{fig:wd_relation}
\end{figure}

\begin{table}[t]
\centering
\caption{Best-fitting parameters for the evaporation correction model.}
\label{tab:fit_parameters}
\begin{tabular}{|c|c|}
\hline
Parameter & Value \\
\hline
$a$ & $6.9267\times 10^7$ \\
$b$ & 9.8273 \\
$c$ & 999.9998 \\
$d$ & 3.2140 \\
$e$ & 1.3835 \\
$g$ & 17.9653 \\
$h$ & 0.6733 \\
$k$ & 0.0014 \\
$n$ & 0.5186 \\
$p$ & -4.3287 \\
\hline
\end{tabular}
\end{table}
In the previous sections, we have shown that evaporation can modify the detachment and subsequent evolution of Huntsman systems. Here we further explore its impact on the final properties of the binaries, in particular on the well-known white dwarf mass ($M_{\rm WD}$)-orbital period relation of NS$+$He WD binaries.

Figure~\ref{fig:wd_relation} presents the simulated $M_{\rm WD}-P_{\rm orb}$ distributions for different evaporation efficiencies, together with two commonly used theoretical relations \citep{2011ApJ...732...70L,2023MNRAS.525.2605G} and observational data. The observational sample of NS$+$He WD binaries is taken from \citet{2023MNRAS.525.2605G}. The simulations reveal a systematic deviation from the standard relation when evaporation is included. In particular, at the low-mass end ($M_{\rm{WD}} \lesssim 0.27\ M_\odot$), systems with strong evaporation ($f=0.1$) exhibit a clear upward shift toward longer orbital periods compared to the non-evaporating baseline, forming an elevated branch in the $M_{\rm{WD}}-P_{\rm{orb}}$ plane. A noticeable dispersion is also present around $M_{\rm{WD}} \approx 0.2-0.27\ M_\odot$, which may be related to structural transitions in the donor during the late stages of MT. In contrast, for $M_{\rm{WD}} \gtrsim 0.27\ M_\odot$, the evolutionary tracks gradually converge toward the standard relation. The overall distribution trend resembles the results presented in \citet{2015ApJ...814...74J} and \citet{2021MNRAS.506.3323T}. To quantify this behavior, we adopt a simple parameterized correction to the baseline $M_{\rm{WD}}-P_{\rm{orb}}$ relation. Specifically, we build on the analytic form proposed by \citet{2011ApJ...732...70L} and introduce an additional evaporation-dependent term,
\begin{equation}
  P_{\rm{orb}}(M_{\rm WD},f)=\left(1 + kf^{n}M_{\rm WD}^{p}\right)P_{\rm{base}}(M_{\rm WD}) \label{eq:relation},
\end{equation}
where $P_{\rm{base}}(M_{\rm WD})$ is given by: 
\begin{equation}
  P_{\rm{base}}(M_{\rm WD})=\frac{aM_{\rm WD}^{b}}{\left(1+cM_{\rm WD}^{d}+eM_{\rm WD}^{g}\right)^{h}}.
\end{equation}
The best-fitting parameters are listed in Table~\ref{tab:fit_parameters}. The fit yields a negative index ($p < 0$), indicating that lower-mass systems are more sensitive to evaporation effects. 

This prescription captures the main trend seen in the simulations, particularly the nonlinear lifting of the low-mass branch. Physically, this behavior can be understood as a consequence of the extended and weakly bound envelopes of low-mass donors. Evaporation-driven mass loss accelerates the removal of the donor envelope and can cause the system to detach at an earlier evolutionary stage, leaving the binary at a wider orbital separation for a given core mass. The effect becomes particularly pronounced around $M_{\rm{WD}} \approx 0.2-0.27\ M_\odot$, where the largest deviation from the standard relation is found. Interestingly, this mass interval partially overlaps with the regime associated with the red-giant branch bump. Structural readjustments occurring during this evolutionary phase may contribute to the increased dispersion and orbital-period elevation seen in the simulations. For higher-mass donors, the envelope becomes more compact and tightly bound. Consequently, the donor is less susceptible to evaporation-driven mass loss, and the impact of evaporation diminishes, leaving the standard $M_{\rm WD}-P_{\rm orb}$ relation largely preserved.

These results show that evaporation can modify the canonical $M_{\rm WD}-P_{\rm orb}$ relation in our evolutionary models, particularly at the low-mass end. However, the current observational data do not provide clear evidence for such evaporation-induced deviations, as the observed systems are generally consistent with the theoretical relations within their measurement uncertainties. Therefore, the current Huntsman observations do not require strong evaporation, but rather provide a useful observational test of this model prediction. Interestingly, the mass range showing the largest model dispersion broadly overlaps with that occupied by several eccentric MSP$+$He WD systems \citep{2022A&A...665A..53S}, although whether the two phenomena are physically connected remains unclear.

\subsection{Summary}\label{s:summary}

In this work, we have investigated the influence of MSP-driven evaporation on the formation and evolution of Huntsman systems by incorporating evaporation into detailed binary evolution calculations. We find that the formation of Huntsman systems is primarily governed by the intrinsic binary evolution, in particular the red-bump-induced detachment during Case B mass transfer. Only systems with initial donor masses in the range $M_{\rm d,0} \simeq 1.0-2.5\ M_\odot$ and orbital periods above the bifurcation period are able to evolve through a Huntsman phase. In this sense, evaporation is not required to trigger the detachment itself, but instead acts on top of the standard evolutionary pathway. Once the system enters the detached phase, evaporation can modify the subsequent evolution in a measurable but generally modest way. It tends to smooth the orbital evolution, slightly affect the duration of the detached phase, and, in cases of sufficiently strong evaporation, may prevent the donor from refilling its Roche lobe. However, its impact on the overall parameter space of Huntsman formation remains limited.

Consistent with this picture, our population synthesis calculations indicate that the number of observable Huntsman systems is $\sim 10$, broadly consistent with the currently known sample. Increasing the evaporation efficiency leads to a slight reduction in the predicted number of systems, while the distribution in the observable parameter space becomes somewhat broader.

We further show that evaporation can modify the canonical $M_{\rm{WD}}-P_{\rm{orb}}$ relation of NS-helium white dwarf binaries, particularly at the low-mass end where it leads to systematically wider orbits in our models. However, the current observations do not provide clear evidence for such evaporation-induced deviations.

Overall, these results suggest that evaporation is not the dominant mechanism in the formation of Huntsman systems, but can modestly modify their subsequent evolution and some properties of their descendants. The current observations do not require strong evaporation, but they provide a useful test of the evaporation-induced effects predicted by our models. Future observations of detached MSP binaries and improved modeling of pulsar-driven mass loss will be essential to further constrain these effects.

\begin{acknowledgements}
We are grateful to the referee for their helpful comments. This work was supported by the startup research fund of Taiyuan University of Technology (Grant No.RY2500004740) and the Natural Science Foundation of China under grant Nos.12041301, 12121003, 123B2045, 12503038 and 12403035.
\end{acknowledgements}

\bibliographystyle{aa} 
\bibliography{ref}

\end{document}